\documentstyle[epsfig]{mn}
\def\msun{M_{\odot}}
\def\simlt{\mathrel{\rlap{\lower 3pt\hbox{$\sim$}}\raise 2.0pt\hbox{$<$}}}
\def\simgt{\mathrel{\rlap{\lower 3pt\hbox{$\sim$}} \raise 2.0pt\hbox{$>$}}}
\def\lsim{\mathrel{\rlap{\lower 3pt\hbox{$\sim$}}\raise 2.0pt\hbox{$<$}}}
\def\gsim{\mathrel{\rlap{\lower 3pt\hbox{$\sim$}} \raise 2.0pt\hbox{$>$}}}

\def\Msun{M_{\odot}}

\begin{document}

\title[The Infrared Glow of First Stars]{The Infrared Glow of First Stars}
\author[Salvaterra, Magliocchetti, Ferrara, Schneider]{Ruben Salvaterra$^1$,
Manuela Magliocchetti$^2$, Andrea Ferrara$^2$, Raffaella Schneider$^{3,4}$\\
$1$ Dipartimento di Fisica e Matematica, Universit\'a dell'Insubria, 
Via Valleggio 11, 22100 Como, Italy\\
$2$ SISSA/International School for Advanced Studies,
Via Beirut 4, 34100 Trieste, Italy\\
$3$ Osservatorio Astrofisico di Arcetri, Largo Enrico Fermi 5, 50125 Firenze, 
Italy\\
$4$``Enrico Fermi'' Centre, Via Panisperna 89/A, 00184 Roma, Italy }

\maketitle \vspace {7cm}

\begin{abstract}
Kashlinsky et al. (2005) find a significant cosmic infrared background fluctuation excess on angular scales $\simgt 50$ arcsec that cannot be explained by instrumental noise or local foregrounds. 
The excess has been tentatively attributed to emission from primordial 
very massive (PopIII) stars formed $\le 200$~Myr after the Big Bang. 
Using an evolutionary model motivated by independent
observations and including various feedback processes, we find that  
PopIII stars can contribute  $< 40$\%  of the total background intensity  
($\nu J_\nu \sim 1-2$ nW m$^{-2}$ sr$^{-1}$ in the 0.8-8 $\mu$m range)
produced by all galaxies (hosting both PopIII and PopII stars)  at $z\ge 5$.
The infrared fluctuation excess is instead very precisely accounted
by the clustering signal of galaxies 
at $z\ge 5$, predominantly hosting PopII stars with masses and properties similar to the present ones.
\end{abstract}


\section{Introduction}

Observations of the infrared background provide important
information on the emission of cosmic luminous sources throughout the history of the
Universe. It has been suggested (Santos, Bromm \& Kamionkowski 2003; 
Salvaterra \& Ferrara 2003) that a large
fraction of the measured Near-InfraRed (1-10 $\mu$m) cosmic 
Background (NIRB) arises from redshifted Ly$\alpha$ line photons and nebular
emission produced by the first very massive metal-free stars. 
This hypothesis, however, is very demanding in terms of the required
conversion efficiency of baryons into stars (Madau \& Silk 2005). 
A large NIRB contribution from such stars has more recently been rejected 
by the paucity ($\le 3$) of $z\sim 10$ candidate sources in 
Hubble Space Telescope ultra-deep observations (Salvaterra \& Ferrara 2005). 
Nevertheless, a more modest contribution from very high redshift 
galaxies, whose clustering should leave a distinct signature on small-scale 
angular fluctuations of the background light (Magliocchetti, Salvaterra \&
Ferrara 2003; Kashlinsky et al. 2004; Cooray et al. 2004), is still possible.

Kashlinsky et al. (2005) have recently found significant NIRB 
fluctuations in deep exposure data obtained with Spitzer/IRAC (Fazio et al. 
2004a, 2004b) in 
four channels (3.6, 4.5, 5.8, and 8 $\mu$m), after Galactic stars and galaxies bright enough to 
be individually
resolved by the instrument have been carefully subtracted. With the only
exception of the 8 $\mu$m channel, the 
shape and amplitude of the power spectrum cannot be reproduced by
either contributions from intervening dusty, Galactic neutral hydrogen
gas (cirrus) or from local interplanetary dust (zodiacal light).
Ordinary galaxies ($z \lsim 5$) produce fluctuations due to their
clustering and shot-noise. The faint flux limits ($\geq 0.3$ $\mu$Jy)
of Spitzer data allow to push their residual clustering contribution
below the level of the excess signal at relatively large ($\gsim 50$~arcsec) 
angular scales (Kashlinsky et al. 2005). 
The shot noise component, estimated directly
from galaxy counts, fits the observed fluctuations at
smaller angular scales, and rapidly fades away at larger angles. 
The residual large scale
signal has been ascribed by Kashlinsky et al. (2005) as coming 
from very distant ($z\ge 5$) sources provided their 
total flux contribution is $> 1$ nW m$^{-2}$ sr$^{-1}$.
The aim of this Letter is to show that this is indeed the case.
 
The layout of the paper is as follows: in Section 2 we will briefly describe the adopted 
model, while in Section 3 we provide predictions for the NIRB intensity and 
fluctuations and compare the latter ones with the results of  Kashlinsky et al. 
(2005). Section 4 summarizes our conclusions.

\section{The Model}

Schneider et al. (2005) have presented a scenario for the formation of galaxies
in a concordance $\Lambda$CDM cosmological model\footnote{We adopt a
$\Lambda$CDM cosmological model with parameters
$\Omega_M=0.3$, $\Omega_\Lambda=0.7$, $h=0.7$, $\Omega_B=0.04$, $n=1$, and
$\sigma_8=0.9$, which lies within the experimental errorbars of WMAP
experiment (Spergel et al. 2003)} which includes a self-consistent
treatment of two key feedback processes:
(i) {\it radiative feedback}, 
suppressing star formation in H$_2$-cooling halos and the formation of low-mass 
galaxies due to the effects of UV background radiation approaching the
reionization epoch, and (ii) {\it chemical feedback}, which controls the transition from metal-free
stars (PopIII) to ordinary stars (PopII) through the progressive enrichment 
of star forming gas with heavy elements released by supernova 
explosions (Schneider et al. 2002, 2004, 2005; Bromm et al. 2001). Chemical
feedback propagates through the hierarchy of galaxy mergers from 
progenitors to their descendants so that, at each redshift, existing halos 
which are allowed to form stars are classified as PopII (PopIII) galaxies 
depending on whether the halo itself or any of its progenitors have (have not) 
already experienced an episode of star formation. 

Within this model we can compute the comoving specific emissivity, $\epsilon_\nu$, 
which is given by 

\begin{equation}
\epsilon_\nu(z)=\int_z^{\infty} dz^\prime l_\nu(t_{z,z^\prime}) 
\int^{M_{max}(z^\prime)}_{M_{min}(z^\prime)} M_\star \frac{d^2 n}{dM_h dz^\prime}(M_h,z^\prime) dM_h,
\label{eq:ep}
\end{equation}

\noindent
where $d^2n/dM_h dz$ is the formation rate of halos of total (dark+baryonic) mass $M_h$ with
corresponding stellar mass $M_\star$,
and $M_{min}(z)$ is the minimum mass of halos (corresponding to a virial temperature of $10^4$~K, 
i.e. $M_{min}(z)\sim 10^8\;\Msun\;(1+z/10)^{-3/2}$) 
allowed to form 
stars at redshift $z$; $M_{max}(z)$ is the maximum mass of halos which depends
on the details of their merging history (see Schneider et al. 2005);
$l_\nu(t_{z,z^\prime})$ is the template specific luminosity for a
stellar population of age $t_{z,z^\prime}$ (time elapsed between redshift 
$z^\prime$ and $z$). 
Following the results of Schneider et al. 2005, both PopII and PopIII stars are
assumed to form according to a Salpeter Initial Mass 
Function (IMF) with an efficiency $f_\star = 0.1$. 
The emission properties of PopII stars are taken from the GALAXEV
library (masses in the range 0.1-100 $\msun$,
metallicity $Z=10^{-2}Z_\odot$; Bruzual \& Charlot 2003) 
and those of PopIII stars are based on the synthesis model for metal-free stars of 
Schaerer (1-100 $\msun$, $Z=0$; Schaerer 2002).

A set of independent observational constraints can be accommodated within
this model (Schneider et al. 2005): the number counts of 
dropout galaxies at $z\sim 6$ ($i$-dropouts; Bouwens et al. 2005b) and at 
$z\sim 10$ ($J$-dropouts; Bouwens et al. 2005a), and the
value of the optical depth for Thomson scattering $\tau_e=0.16\pm0.04$ 
measured by the WMAP satellite (Kogut et al. 2003). 
Indeed, the predicted surface density of $i$-dropouts  at a limiting
magnitude of $i=28$ is $\sim5$ arcmin$^{-2}$ (observed value 
$\sim 4.7$ arcmin$^{-2}$; Bouwens et al. 2005b). 
The model is also consistent with the three
candidate dropouts at $z\sim 10$ reported by Bouwens 
et al. (2005a). Moreover, the large value of $\tau_e$ can be 
matched with standard values of the ionizing photon escape fraction from
galaxies ($f_{esc} \leq 0.2$) and gas clumping factor $C \sim 10$,
resulting in a reionization redshift of 13.2 (Schneider et al. 2005).

\section{NIRB intensity and fluctuations}

The model presented in Section 2 (referred to as model C in Schneider et al. 2005) allows 
to compute  the NIRB intensity $J_{\nu_{0}}$ from $z\ge 5$ galaxies seen at frequency 
$\nu_{0}$ by an observer at redshift $z_{0}=0$. This can be done in a fashion similar to 
Salvaterra \& Ferrara (2003) as:

\begin{equation}
J_{\nu_{0}} (z_0)= \frac{(1+z_0)^3}{4\pi}\int^{\infty}_{z_{0}}
\epsilon_\nu(z)e^{-\tau_{eff}(\nu_{0},z_{0},z)}\frac{dl}{dz}dz,
\end{equation}

\noindent
where $\nu=\nu_0(1+z)/(1+z_0)$, $dl/dz$ is the proper line element, $\tau_{eff}(\nu_{0},z_{0},z)$ 
is the effective optical depth at $\nu_0$ of the intergalactic medium between 
redshift $z_0$ and $z$ (see Section 2.2 of Salvaterra \& Ferrara 2003 for a full
description of the IGM modelling), and $\epsilon_\nu(z)$ is provided by equation (1).
Our results are shown in Figure 1. The background intensity, $\nu J_\nu \sim 1-2$ 
nW m$^{-2}$ sr$^{-1}$, is almost constant in the 0.8-8 $\mu$m range. 
PopII galaxies dominate the NIRB in the entire wavelength 
range, while PopIII galaxies contribute at most 40\% of the total intensity
(at $\lambda\sim 1.5\;\mu$m), via their strong Ly$\alpha$ line emission. 
We have checked that the sources responsible for such emission are too
faint to be resolved and identified at the Spitzer magnitude limit in all
four channels. 
The total contribution of all galaxies at $z\ge 5$ is 
1.55, 1.45, 1.07 and 0.74 nW m$^{-2}$ sr$^{-1}$ in the 3.6, 4.5, 5.8 and 8.0 IRAC bands, 
respectively, representing a substantial fraction ($\sim$ 10-20\%) of the total NIRB intensity
estimated from integration of Spitzer galaxy number counts (Fazio et al. 2004b).

\begin{figure}
\begin{center}
\centerline{\psfig{figure=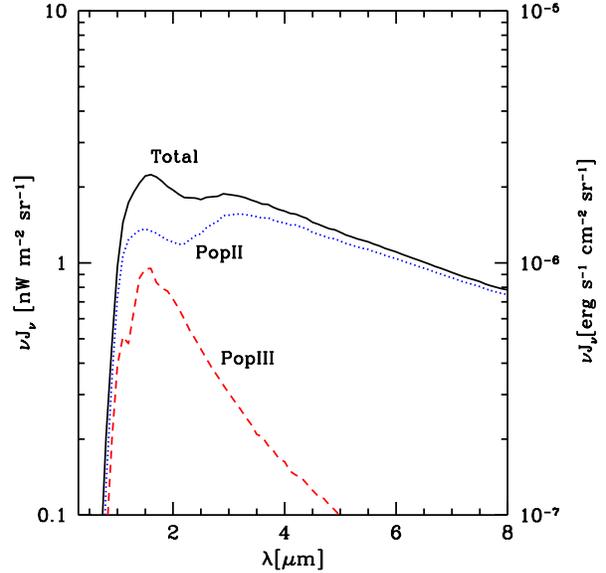,height=8cm}}
\caption{IR background from sources 
at $z\ge 5$ in the wavelength range 0.3-8 $\mu$m. The dotted (dashed) line shows the contribution of PopII 
(PopIII) galaxies. The sum of the two components is denoted by the solid line.
The sharp drop below $0.8$ $\mu$m is due to the absorption of the 
intergalactic medium for $\lambda<\lambda_{{\rm Ly}\alpha}$ in the emitter
rest frame.}
\label{fig1}
\end{center}
\end{figure}

The angular correlation function of intensity fluctuations $\delta J$ due to
inhomogeneities in the space distribution of unresolved sources
(i.e. with fluxes fainter than some threshold $S_d$) is defined as:
\begin{eqnarray}
C(\theta)=\langle \delta J(\theta^\prime, \phi^\prime)\; \delta
J(\theta'', \phi'')\rangle ,
\label{eq:ctheta}
\end{eqnarray}
where $(\theta^\prime, \phi^\prime)$ and $(\theta'', \phi'')$ identify
two positions on the sky separated by an angle $\theta$. The above expression
can be written as the sum of two terms, $C_P$ and $C_C$, the first one due
to Poisson noise (i.e. fluctuations given by randomly distributed
objects), and the second one owing to source clustering. It can be shown that the 
shot noise contribution originating from $z\geq 5$ galaxies
is negligible, so we only concentrate on fluctuations which stem from 
the clustering of these sources, i.e. we assume $C(\theta)\equiv C_C$.

The method adopted here is similar to that presented in Magliocchetti et al. (2003), whereby angular 
fluctuations are obtained by means of the expression:

\begin{eqnarray}
C(\theta)=\left({1\over 4\pi}\right)^2  \int_{z_0}^{\infty}\!\!\!\!\!\!dz\;
\frac{\epsilon_{\nu}^2(z)} {(1+z)^2}\;e^{-2\tau_{eff}}\left(\frac{dx}{dz}\right)
\nonumber \\
\times \int_{-\infty}^\infty du\;\xi_g(r,z),
\label{eq:cth}
\end{eqnarray}
where $x=l(1+z)$ is the comoving coordinate, $r=(u^2+x^2\theta^2)^{1/2}$
(for a flat universe and in the small angle approximation), and $\epsilon_\nu$ is 
defined as in Section 2. 

The spatial two-point correlation function of a class of galaxies '$g$',
$\xi_g(r,z)$, in general results from a complicated interplay between the
clustering properties of the underlying dark matter and physical processes
associated to the formation of such galaxies (see e.g. 
Magliocchetti \& Porciani 2003). However, the sources we are considering 
in this work are small enough (typical dark matter masses 
around $10^9-10^{10} M_{\odot}$ for $z\simeq 5$, values which rapidly decrease when 
moving to higher redshifts, see Schneider et al. 2005 for further details) and set at 
high enough redshifts to ensure that -- if present --
their sub-halo behavior falls into angular scales which are too small to be detected.  
We can therefore safely assume a one-to-one correspondence between halos and galaxies 
and write: $\xi_g(r,z)=\xi(r,z)b^2_{\rm eff}(z)$, 
where $\xi(r,z)$ is the mass-mass correlation function (evaluated by following 
the Peackock \& Dodds 1996 approach up to 100 Mpc comoving) and $b_{\rm eff}
(z)$ the bias associated to all dark matter halos massive enough
to host a galaxy at redshift $z$. This latter quantity was then obtained respectively 
for PopII and PopIII galaxies
by integrating the function $b(M_h,z)$ - representing the bias of
individual haloes of mass $M_h$ - opportunely weighted by the number
density of such sources $dn/dM_h (M_h,z)\; f(M_h,z)$ as:
\begin{eqnarray}
b_{\rm eff}(z)=\frac{\int_{M_{ min}(z)}^{M_{max}(z)} dM_h\;b(M_h,z)\;\frac{dn}{dM_h}(M_h,z)\;f(M_h,z)}
{\int_{M_ {min}(z)}^{M_{max}(z)}dM_h\;\frac{dn}{dM_h}(M_h,z)\;
f(M_h,z)},
\label{eq:beff}
\end{eqnarray}
where $dn/dM_h(M_h,z)$ is the Sheth \& Tormen (1999) mass function of the dark matter halos,  
$f(M_h,z)$ is the fraction of halos that at any mass and redshift host 
a PopII (PopIII) galaxy and $M_{min}(z)$, $M_{max}(z)$ 
are the limiting mass values for such halos to host either a PopII or a PopIII
source (see Section 2). 

The above prescriptions then allow to estimate $C(\theta)$, i.e. the contribution from the 
clustering of unresolved $z\ge 5$ (PopII and PopIII) galaxies to the background fluctuations at 
different wavelengths. An important feature to mention about the resulting $C(\theta)$ is that -- 
independent of the considered frequency and only due to the high redshift of the sources 
in exam -- it features a sharp drop at around 300 arcsecond beyond which it rapidly approaches zero. 
This implies that signals on scales larger than the above value such as those detected by some 
experiments (e.g. Hauser \& Dwek 
2001; Matsumoto et al. 2005) have to rely on the presence of a more local population of unresolved sources.

\begin{figure}
\begin{center}
\centerline{\psfig{figure=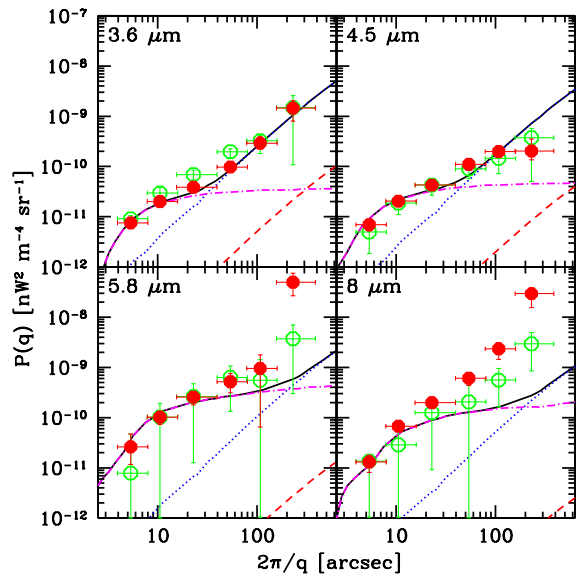,height=8cm}}
\caption{Power spectrum of fluctuations at the IRAC bands.
Different symbols show data of different observed fields (Kashlinsky et al. 2005): 
filled circles corresponds to the QSO 1700 field, open circles to EGS. 
Data are averaged over wide bins to increase the signal-to-noise ratio. The errors
are $N_q^{-1/2}$, where $N_q$ is the 
number of Fourier elements at the given $q$-bin. Dotted (dashed) lines show the 
contribution from PopII (PopIII) galaxies. Dot-dashed lines show the contribution of shot 
noise from remaining galaxies fainter than the limiting magnitude (Kashlinsky et al. 2005).
The solid line is the sum of the different components.}
\label{fig2}
\end{center}
\end{figure}

Finally, we can move to the calculation of the power spectrum $P(q)$ originating from the above 
fluctuations, 
\begin{equation}
P(q)=2 \pi \int_0^\infty C(\theta) J_0(\theta,q) \theta d\theta, 
\end{equation}
where $J_0$ is the zero-th order Bessel function. 

In order to confront our predictions with the Kashlinsky et al. (2005) 
results, $P(q)$ has been evaluated in the 3.6, 4.5, 5.8 and 8 $\mu$m IRAC bands.
Our findings are summarized in Figure 2, where the dotted lines represent the 
fluctuations originating from PopII galaxies, the dashed lines those due to PopIII 
sources and the dot-dashed curves indicate the shot noise contribution 
from galaxies fainter than the limiting magnitude as estimated by Kashlinsky et al. (2005).
It is clear from the plot that at the lowest frequencies the contribution of $z \geq 5$
galaxies can provide an excellent fit to the observed fluctuations; in particular it dominates 
over the shot-noise component at all angular scales greater than $\sim 50$ arcsec. 
At 5.8 $\mu$m, their importance is reduced but the presence of $z\ge 5$ sources is still 
required to properly account for the amplitude of the signal at the largest probed scales. 
Finally, high redshift galaxies and shot-noise alone cannot reproduce the power spectrum at 
8 $\mu$m, where the cirrus probably becomes the dominant component (Kashlinsky et al. 2005).

As a last point it is worth mentioning that, as expected within the present scenario, the contribution of 
PopIII galaxies is negligible in all IRAC bands; the amplitude of their signal is in fact $\sim
50$ times lower than that of PopII galaxies in the most favorable case (3.6 $\mu$m), while 
the PopIII-to-PopII contribution ratio
can go as low as $10^{-3}$ at the highest frequency probed by IRAC.

\section{Conclusions}

Using a physically-motivated, observationally-tested model of the early
Universe (Schneider et al. 2005), we compute the expected background
radiation in the NIR by sources forming when the Universe was $<1$ Gyr old. 
We find that the background intensity, $\nu J_\nu \sim 1-2$ nW m$^{-2}$ 
sr$^{-1}$, is almost constant in the 0.8-8 $\mu$m range. 
PopII galaxies dominate the NIRB in the entire wavelength 
range, while PopIII galaxies contribute at most 40\% of the total intensity
(at $\lambda\sim 1.5\;\mu$m), via their strong Ly$\alpha$ line emission.
Finally, we found that the infrared fluctuation excess on angular 
scales $\geq 50$ arcsec detected by Spitzer/IRAC (Kashlinsky et al. 2005)
is accounted very precisely by the clustering signal of galaxies at 
$z\ge 5$ predominantly hosting stars with masses and properties similar to the 
present ones. 

Two additional points are worth noticing:
(i) very massive stars ($M \geq 100 \msun$) do not need to be invoked
to explain NIRB fluctuations and reionization history; (ii) because of
their small contribution ($P(q) \leq 10^{-10}$ nW$^2$ m$^{-4}$ sr$^{-1}$)
to the observed power spectrum in all channels, extracting the signal of the 
(very) first PopIII stars is extremely challenging.
Future instruments (as the James Webb Space Telescope) 
will be able to directly identify these sources up to $z=10$ or above. 
Finally, the intensity of the NIRB provided by $z\ge 5$ 
galaxies falls short of accounting for the excess
measured by IRTS (Matsumoto et al. 2005) and DIRBE (Hauser \& Dwek 2001)
experiments. The origin of this component remains very puzzling (Salvaterra
\& Ferrara 2005) and might require either a revision of current model of zodiacal
light subtraction or the existence of a large population of faint galaxies
located at $z\simeq 2-3$ (or both). Important insights on these issues are expected from the
upcoming CIBER experiment (Bock et al. 2005), that will be able to simultaneously measure  
the total NIRB intensity and fluctuation power spectrum in the poorly known wavelength 
range 0.8-2 $\mu$m. Such instrument, in addition, will allow a clear separation of the cosmological 
signal from local foregrounds (i.e. zodiacal light).

\end{document}